\documentclass[aps,prb,twocolumn,superscriptaddress,raggedbottom,showpacs,floatfix]{revtex4}
\usepackage{amsmath,amsfonts,amssymb,amsthm}
\usepackage{color}
\usepackage{graphicx,latexsym,cancel}

\def\p{{\rm p}}
\def\pf{{\rm pf}}
\def\e{{\rm e}}
\def\i{{\rm i}}
\def\d{{\rm d}}
\def\ast{\dagger}
\def\fast{{\phantom{\dagger}}}
\newtheorem{theorem}{Theorem}[section]

\begin{document}

\title{Non-equilibrium steady-states for interacting open systems: exact results}

\author{Valeriu Moldoveanu}
\affiliation{National Institute of Materials Physics, P.O. Box MG-7,
Bucharest-Magurele, Romania}
\affiliation{  Department of Mathematical Sciences, Aalborg
     University, Fredrik Bajers Vej 7G, 9220 Aalborg, Denmark }

\author{Horia D. Cornean}
\affiliation{  Department of Mathematical Sciences, Aalborg
     University, Fredrik Bajers Vej 7G, 9220 Aalborg, Denmark }

\author{Claude-Alain Pillet }
\affiliation{Centre de Physique Th\'eorique, CNRS -- Universit\'es de Provence, de la M\'editerran\'ee
et du Sud Toulon-Var, B.P. 20132, 83957 La Garde, France }

\begin{abstract}
Under certain conditions we prove the existence of a steady-state transport regime for interacting
mesoscopic systems coupled to reservoirs (leads). The partitioning and partition-free scenarios
are treated on an equal footing. Our time-dependent scattering approach is {\it exact} and proves,
among other things the independence of the steady-state quantities from the initial state of the 
sample. Closed formulas for the 
steady-state current amenable for perturbative calculations w.r.t.\;the interaction strength are 
also derived. In the partitioning case we calculate the first order correction and recover the 
mean-field (Hartree-Fock) results.
\end{abstract}


\maketitle

\section{ Introduction}

 The theoretical modeling of time-dependent transport has been an active area of research in 
the last few years\cite{Stefanucci,Kurth1,Myohanen}. Transient currents are calculated 
using the Keldysh formalism \cite{JWM}, and the electron-electron interaction (EEI) effects are
accounted for via time-dependent density functional theory (TDDFT) or many-body perturbative 
(MBP) methods. But no matter which metod one uses and how simple the system is, one recovers 
an old and often avoided question in non-equilibrium transport: does there exist a non-equilibrium 
steady-state (NESS), and if yes, is it unique? 

Recently, Kurth {\it et al.} \cite{Kurth2} presented TDDFT simulations for a single-level quantum 
dot (QD) in the Coulomb blockade regime. By selecting a suitable exchange-correlation potential 
they noticed that the system does not evolve to a steady state but rather follows charging/discharging
cycles. Moreover, My\"{o}h\"{a}nen {\it et al.} \cite{Myohanen} and Puig {\it et al.} \cite{Puig} 
emphasized that different approximation schemes for the interaction self-energy lead to different 
steady-states, i.e, in the long-time limit the numerical simulations lead to different 
values of the current. These recent findings cleary show that the crossover to a steady-state (if any) 
is a non-trivial aspect which is revealed only by a fully time-dependent formalism for open and 
interacting systems. We remind here that following Ref. \cite{JWM} the Keldysh formalism was 
extensively used to compute the stationary currents  by assuming i) that such a steady state is 
achieved and ii) that the interaction strength is rather small such that a perturbative approach makes 
sense.  The first assumption implies that it is sufficient to work directly with the Fourier transforms 
of the Green functions. The second assumption allows one to exploit diagrammatic techniques and 
conserving approximations for the interaction self-energy\cite{Thygesen}. Given the results 
mentioned above one faces three questions:
1) How legitimate is it to take for granted the steady-state quantities of the Keldysh formalism 
especially in the presence of electron-electron interaction ? 
2) Is the result of Kurth {\it et al.} \cite{Kurth2} universal? 
3) Is it possible to establish the existence of a stationary current in the long-time limit 
{\it before} selecting a given approximation scheme for the explicit calculations of the interaction 
effects?

In this note we prove that an interacting sample evolves to a NESS
provided: 
i) all single-particle eigenstates of the isolated sample become resonances (with positive width) 
when the leads are coupled and 
ii) the interaction strength is small enough to ensure convergence of a certain perturbation expansion. 
These conditions are met for large quantum dots coupled to broad leads. 
Condition i) is also fulfilled if the bias applied on the leads covers the entire spectrum of the sample 
(wide-band limit). We follow the scattering approach to the NESS of open quantum systems
advocated by Ruelle \cite{Ru} and implemented, in the fermionic case, by Fr\"ohlich et al.\cite{FMU}
and Jak\v si\'c et al. 
\cite{Jaksic5,Jaksic6,Jaksic7}. The existence of a steady-state is rigorously proved by deriving 
explicit expressions for both the lesser Green's function and the current in the infinite time limit.  
Our method is exact in the interaction and needs neither 
Langreth rules, nor Dyson equations for Keldysh-Green's functions. It provides convergent
expansions in terms of the interaction strength, i.e., it shows that the NESS is an analytic
function of the interaction.

Moreover, the proof of the steady-limit covers the two complementary but 
different transport scenarios: the partitioning approach of Caroli {\it et al.} \cite{Caroli} and the 
partition-free setting coined by Cini \cite{Cini}. Let us briefly remind them here. In the partitioning 
approach the central region (sample) is coupled to {\it biased} leads at some initial instant. In contrast,
the partition-free setup starts from a coupled and unbiased system, a bias being switched at $t=t_0$. 
Both settings have also received interest from the mathematical point of view  
\cite{Aschbacher,Nenciu,CNZ,CGZ,Jaksic6}. 
   
The content of the paper is as follows: Section II sets the model and some notations 
needed for the partitioning and partition-free settings. We formulate our main result, the steady
state limit of the lesser Green's function, in Section III.  The expression for the steady currents is 
derived in Section IV, along with the Landauer-B\"{u}ttiker formula which holds if the
Coulomb effects are considered up to the 1st order term in the interaction strength. 
We outline the proof of our results in Section V and conclude in Section VI with some
general comments.

\section{ Model and notations}
Our system consists of a finite sample $S$ coupled to M semi-infinite leads labeled by $\gamma$.
It is described by a discrete model: sites from the lead $\gamma$ are denoted by 
$\{i_{\gamma}\}_{i\geq 0}$ and $\{m\}_{m\in S}$ are the sites of the sample. 
We denote by $\mathcal{H}_S$ and $\mathcal{H}_L$
the one particle Hilbert spaces of the sample and leads. The one particle Hilbert space of the 
compound system is $\mathcal{H}=\mathcal{H}_L\oplus\mathcal{H}_S$ and $\mathcal{F}$ denotes 
the fermionic Fock space over $\mathcal{H}$. The one particle Hamiltonian of the noninteracting 
sample is $h_S$, an arbitrary self-adjoint operator on $\mathcal{H}_S$. In terms of the on-site 
creation/annihilation operators on $\mathcal{F}$, the Hamiltonian and number operator of the
sample are
$$
H_S=\sum_{m,n\in S}\langle m|h_S|n\rangle\, a^\ast_m a^\fast_{n},\qquad
N_S=\sum_{m\in S}a^\ast_m a^\fast_{m}.
$$
The semi-infinite leads are described by the Hamiltonian $H_L=\sum_\gamma H_\gamma$ and 
number operator $N_L=\sum_\gamma N_\gamma$, where
$$
H_\gamma=\tau_L\sum_{i\ge0} \left( a^\ast_{i_\gamma}a^\fast_{(i+1)_\gamma}
\!\!+ a^\ast_{(i+1)_\gamma}a^\fast_{i_\gamma}\right),\ \;
N_\gamma=\sum_{i\ge0} a^\ast_{i_\gamma}a^\fast_{i_\gamma}.
$$
We also need the tunneling Hamiltonian $H_T$, some constant potential $H_B$ applied to the leads 
and the electron-electron interaction in the sample $V$,
\begin{multline*}
H_T=\tau\sum_{\gamma=1}^M
\left (a^\ast_{0_\gamma}a^\fast_{m_\gamma}\!\!+a^\ast_{m_\gamma}a^\fast_{0_\gamma}\right ),
\qquad
H_B=\sum_{\gamma=1}^Mv_{\gamma}N_\gamma,\\
\hfill V=\frac{\xi}{2}\sum_{m,n\in S}v(m,n)\,a^\ast_m a^\fast_m a^\ast_{n} a^\fast_{n}.\hfill
\end{multline*}
Here, $\tau$ is the coupling strength, $m_{\gamma}\in S$ the contact site of the sample with lead 
$\gamma$, $v_{\gamma}$ a constant potential, $\xi$ the interaction strength and $v(m,n)$ 
a pair potential. The differences $v_{\alpha}-v_{\beta}$ define the bias between the corresponding 
leads. By convention all the $v_\gamma$ vanish in the partitioning case, so that $H_B=0$.
For both partitioning (`$\p$')/partition-free (`$\pf$') settings we assume that the switching 
(of the coupling to the leads $H_T$/of the bias $H_B$) happens suddenly at $t=0$. A smooth 
switching can also be treated up to some technicalities \cite{CMP}, and does not influence the results.

The Hamiltonians $H_S,H_L,H_T,H_B$ as well as
\begin{eqnarray*}
H_0=H_S+H_L+H_B, &\quad& H=H_0+H_T,
\end{eqnarray*}
act in the  Fock space ${\cal F}$ as the second quantized versions of single particle tight-binding 
Hamiltonians $h_S,h_L,h_T,h_B,h_0,h$ acting on ${\cal H}$. 
Similar relations hold for the number operators $N_S,N_\gamma,N_L$ and 
$N=N_S+N_L$. We denote by $\mathcal{F}_L$ and $\mathcal{F}_S$ the subspaces of 
$\cal F$ where $N_S=0$ and $N_L=0$ respectively.
The full dynamics of the system is generated by the Hamiltonian
$$
K=H+V.
$$

We now introduce the initial state of the system as thermodynamic limit of
states defined by density matrices on the sample coupled to finite leads
of length $\Lambda$. Indicating this infrared cutoff by the superscript ${}^{(\Lambda)}$,
we set
$$
\rho_{L,\vec\mu}^{(\Lambda)}=
\frac{\e^{-\beta(H_L^{(\Lambda)}-\sum_\gamma\mu_\gamma N_\gamma^{(\Lambda)} )}}
{{\rm Tr}_{{\cal F}_L^{(\Lambda)}}\{\e^{-\beta(H_L^{(\Lambda)}
-\sum_\gamma\mu_\gamma N_\gamma^{(\Lambda)} )}\}},
$$
for $\vec\mu=[\mu_1,...,\mu_M]$ and, for any leads observable $\mathcal{O}_L$,
$$
\langle\mathcal{O}_L\rangle_{L,\vec\mu}
=\lim_{\Lambda\to\infty}\mathrm{Tr}_{\mathcal{F}_L}\{\rho_{L,\vec\mu}^{(\Lambda)}\mathcal{O}_L\}.
$$
We recall that this state is characterized by the two-point function
$$
\langle a^\ast_{j_\alpha} a^\fast_{i_\gamma}\rangle_{L,\vec\mu}=\delta_{\alpha\gamma}
\langle i_\gamma|F_\gamma(h_\gamma)|j_{\gamma}\rangle,
$$
and Wick's theorem. There, $F_\gamma(\varepsilon)=
(1+\e^{\beta(\varepsilon-\mu_\gamma)})^{-1}$ denotes the Fermi-Dirac function of lead $\gamma$.

In the `$\p$' setting, the initial state is the product state defined by
$$
\langle\mathcal{O}_L\mathcal{O}_S\rangle_\p=
\langle\mathcal{O}_L\rangle_{L,\p}
\,\langle\mathcal{O}_S\rangle_{S,\p}
$$
for any observables $\mathcal{O}_{L/S}$ of the leads/sample. There,
$$
\langle\mathcal{O}_L\rangle_{L,\p}=\langle\mathcal{O}_L\rangle_{L,\vec\mu},\qquad
\langle\mathcal{O}_S\rangle_{S,\p}=\mathrm{Tr}_{\mathcal{F}_S}\{\rho_{S,\p}\mathcal{O}_S\},
$$
where $\rho_{S,\p}$ is an arbitrary density matrix on $\mathcal{F}_S$.
In the `$\pf$' case the leads and sample are coupled and have the same 
chemical potential $\mu_0$, i.e., 
$$
\langle\mathcal{O}\rangle_\pf=
\lim_{\Lambda\to\infty}\mathrm{Tr}_{\mathcal{F}^{(\Lambda)}}\{\rho^{(\Lambda)}_\pf\mathcal{O}\},
$$
where
$$
\rho_{\pf}^{(\Lambda)}=\frac{\e^{-\beta(H_S+H_L^{(\Lambda)}+H_T+V-\mu_0 N^{(\Lambda)})}}
{{\rm Tr}_{\mathcal{F}^{(\Lambda)}} \{\e^{-\beta(H_S+H_L^{(\Lambda)}+H_T+V-\mu_0 N^{(\Lambda)})} \}}.
$$
For later reference, we also define the leads state
$$
\langle\,\cdot\,\rangle_{L,\pf}=\langle\,\cdot\,\rangle_{L,\vec\mu=[\mu_0,\ldots,\mu_0]}.
$$

The {\sl lesser Green's function} is defined as
$$
G^{<}_{\boxdot\,xy}(t,s)=\i\,\langle a^\ast_y(s)a^\fast_x(t)\rangle_\boxdot,
$$
where $x,y$ are sites from either leads or sample,
$a^\fast_x(t)=\e^{\i tK}a^\fast_x\e^{-\i tK}$ and $\boxdot$ stands for either `$\p$' or `$\pf$'. 
This object plays a central role in the Keldysh approach and allows to compute
both the particle density and the currents. Our main
concern being the existence of steady currents, we are primarily interested
in its large time behaviour $t,s\to\infty$ for constant $t-s$. 

\section{The existence of the steady-state}
For $|\phi\rangle\in\mathcal{H}$, we set 
$$
a(|\phi\rangle)=\sum_x\langle \phi|x\rangle\,a_x,\quad
a^\ast(|\phi\rangle)=\sum_x\langle x|\phi\rangle\,a_x^\ast,
$$
and let $a^\#(|\phi\rangle)$ denote either $a(|\phi\rangle)$ or $a^\ast(|\phi\rangle)$.
If $q$ is a self-adjoint operator on $\cal H$ and $Q$ denotes
its second quantized version, then the well known identity
\begin{equation}
\e^{\i Q}a^{\#}(|\phi\rangle)\e^{-\i Q}=a^{\#}(\e^{\i q}|\phi\rangle),
\label{wellknown}
\end{equation}
holds. The interaction picture operators
\begin{equation}
\mathcal{A}^\#_x(t)=\e^{-\i tH}a^\#_x(t)\e^{\i tH},
\label{calAdef}
\end{equation}
satisfies the equation of motion
\begin{equation}
\mathcal{A}^\#_x(t)=a^\#_x
+\i\int_0^t[\e^{-\i uH}V\e^{\i uH},\mathcal{A}^\#_x(u)]\d u.
\label{EOM}
\end{equation}
By Eq.\eqref{wellknown} we have
\begin{equation}
\label{Vintrot}
\begin{split}
&\e^{-\i uH}V\e^{\i uH}=\frac{\xi}{2}\sum_{m,n\in S}v(m,n)\\[5pt]
\times &a^\ast(\e^{-\i uh}|m\rangle)a(\e^{-\i uh}|m\rangle)
a^\ast(\e^{-\i uh}|n\rangle)a(\e^{-\i uh}|n\rangle).
\end{split}
\end{equation}
Thus, the Dyson expansion of $\mathcal{A}^\#_x(t)$ obtained by iteration of Eq.\eqref{EOM}
is a sum of iterated integrals involving monomials of the type
\begin{equation}
\label{LemmaL1}
\mathcal{M}(|\phi_1\rangle,\ldots,|\phi_k\rangle)=a^{\#_1}(|\phi_1\rangle)\cdots a^{\#_k}(|\phi_k\rangle),
\end{equation} 
each $|\phi_j\rangle$ being either $|x\rangle$ or $\e^{-\i uh}|m\rangle$ for some
$m\in S$ and $u\in[0,t]$ (so that, in particular $|\langle\Psi|\mathcal{M}|\Phi\rangle|\le1$
for any unit vector $|\Psi\rangle,|\Phi\rangle\in\mathcal{F}$). 
Moreover, one easily sees that this expansion converges for
{\sl any} $t$. In fact, a careful study of this expansion shows that, under suitable
assumptions, it remains convergent even for $t=\infty$.
\begin{theorem}\label{Asymptot}
Assume that the single particle Hamiltonian $h$ has neither eigenvalue nor real resonance. 
If the interaction strength $\xi$ is small enough, then the limits
$$
\mathcal{A}^\#_x=\lim_{t\to\infty}\mathcal{A}^\#_x(t),
$$
exist. Moreover, a convergent expansion of $\mathcal{A}^\#_x$ in powers of the interaction strength
$\xi$ is obtained by setting $t=\infty$ in the Dyson expansion of $\mathcal{A}^\#_x(t)$.
\end{theorem}

{\noindent\bf Remark.}
The first hypothesis of the previous theorem requires some comments. The spectrum of
$h_L+h_B$ is continuous, filling the union of $[v_\gamma-2\tau_L,v_\gamma+2\tau_L]$.
If all the eigenvalues of the isolated sample $h_S$ are embedded in these bands
and if the coupling to the leads $\tau$ is weak enough then all these eigenvalues will generically 
turn into resonances of positive width. In such circumstances, the spectrum of
$h$ is continuous and coincides with that of $h_L+h_B$. Moreover, one can show that
for any $x,y$ in either the leads or the sample,
\begin{equation}
\int_0^\infty|\langle x|\e^{-\i th}|y\rangle|\,\d t<\infty.
\label{decay}
\end{equation}
However, as the coupling $\tau$ increases, some resonances may become real,
cross a band boundary and turn into an eigenvalue of $h$, invalidating \eqref{decay}.
The first hypothesis in Theorem \ref{Asymptot} is meant to ensure the validity of Eq.\eqref{decay}.

\medskip
Let $w$ be an operator on $\mathcal{H}$ such that $|\langle\phi|w|\psi\rangle|\le1$ 
for all unit vectors $|\phi\rangle,|\psi\rangle\in\mathcal{H}$.
Replacing each term \eqref{LemmaL1} in the Dyson expansion of $\mathcal{A}^\#_x$ by 
$\mathcal{M}(w|\phi_1\rangle,\ldots,w|\phi_k\rangle)$ does not alter the convergence
of this expansion. We denote by $\mathcal{A}^\#_x[w]$ the operator obtained
from this modified Dyson expansion.

We are now in position to state our main result (recall that $h_B=0$ in the
`$\p$' case):
\begin{theorem}\label{MainRes}
Under the assumptions of Theorem \ref{Asymptot} one has, for any $s$,
\begin{equation}\label{lesserSS}
\begin{split}
\mathcal{G}^{<}_{\boxdot xy}(s)&=\lim_{t\to\infty}G^{<}_{\boxdot\,xy}(t,t-s)\\
&=\i\langle\mathcal{A}^\ast_y[\e^{-\i s(h_L+h_B)}\omega_+^\ast]
\mathcal{A}^\fast_x[\omega_+^\ast]\rangle_{{L,\boxdot}}.
\end{split}
\end{equation}
There, $\omega_+$ denotes the M\o ller operator \cite{Taylor}
\begin{equation}\label{Moller1}
 \omega_+|\phi\rangle=\lim_{t\to-\infty}\e^{\i th}\e^{-\i th_0}p_L|\phi\rangle,
\end{equation}
where $p_L$ projects on the leads subspace ${\cal H}_L$.
\end{theorem}
In Section \ref{proof}, we shall outline the proofs of Theorems \ref{Asymptot}, \ref{MainRes}. 
Complete mathematical details will be given elsewhere \cite{CMP}. We conclude this section
with several remarks.

\vspace{0.2cm}
\noindent1. Asymptotic completeness\cite{Taylor} implies that
\begin{equation}
\begin{split}
\omega_+^\ast|\psi\rangle
&=\lim_{t\to-\infty}\e^{\i th_0}\e^{-\i th}|\psi\rangle\\
&=\lim_{t\to-\infty}p_L\e^{\i t(h_L+h_B)}\e^{-\i th}|\psi\rangle,
\end{split}
\label{omegastar}
\end{equation}
is unitary from $\mathcal{H}$ to 
$\mathcal{H}_L$ so that the object under the expectation on the RHS of Eq.\eqref{lesserSS} 
has a convergent expansion as described above. Moreover, the expectation is w.r.t.\;the
leads state  $\langle\,\cdot\,\rangle_{{L,\boxdot}}$ which does not depend on the interaction $V$,
i.e., satisfies Wick's theorem.

\noindent 2. Eq.\eqref{lesserSS} implies right away that the expected particle number
in the sample reaches a steady value in the long-time limit
$$
\lim_{t\to\infty}\langle N_S(t)\rangle_\boxdot=-\i\sum_{m\in S}\mathcal{G}^{<}_{\boxdot,mm}(0).
$$
In fact, one can show that, under the assumptions of Theorem \ref{Asymptot},
the system reaches a NESS $\langle\,\cdot\,\rangle_{\boxdot+}$ described by
\begin{align*}
\langle a_{x_1}^{\#_1}\cdots a_{x_k}^{\#_k}\rangle_{\boxdot+}&=
\lim_{t\to\infty}\langle a_{x_1}^{\#_1}(t)\cdots a_{x_k}^{\#_k}(t)\rangle_\boxdot\\
&=\langle \mathcal{A}_{x_1}^{\#_1}[\omega_+^\ast]\cdots
\mathcal{A}_{x_k}^{\#_k}[\omega_+^\ast]\rangle_{L,\boxdot}.
\end{align*}
In the `$\p$' case, this NESS is {\it independent} on
the initial state of the sample  $\rho_{S,\p}$. Moreover, in the special case
$\vec\mu=[\mu_0,\ldots,\mu_0]$,  $\langle\,\cdot\,\rangle_{\p+}$
is the unique equilibrium state of the interacting system at inverse temperature
$\beta$ and chemical potential $\mu_0$.

\noindent 3. In the non-interacting case ($V=0$), the NESS satisfies Wick's theorem with
the two points function $\langle a^\ast_y a^\fast_x\rangle_{\boxdot+}=\langle x|\rho_+|y\rangle$,
where the one-particle density operator $\rho_+$ is given by \cite{Aschbacher, Nenciu}
\begin{equation}
\rho_+=\omega_+\left(\bigoplus_\gamma F_{\gamma}(h_\gamma)\right)\omega_+^\ast.
\label{rhos}
\end{equation}

\section{The steady-state current}

The current operator of lead $\alpha$
$$
J_\alpha=-\left.\frac{\d\ }{\d t}N_\alpha(t)\right|_{t=0}=-\i[K,N_\alpha]
=-\i[H,N_\alpha],
$$
is the second quantized version of the single-particle current $j_{\alpha}=-\i[h,p_\alpha]$, where
$p_\alpha$ projects on lead $\alpha$. Its statistical average is \cite{JWM}
\begin{equation}\label{current1}
\langle J_{\alpha}(t) \rangle_\boxdot
=\tau(G^{<}_{\boxdot m_{\alpha}0_{\alpha}}(t,t)
-G^{<}_{\boxdot 0_{\alpha}m_{\alpha}}(t,t)).
\end{equation}
We introduce the interaction picture current operator
\begin{equation}
\label{Bcal}
\mathcal{J}_\alpha(t)=\e^{-\i tH}\e^{\i tK}J_{\alpha}\e^{-\i tK}\e^{\i tH},
\end{equation}
which is similar to ${\cal A}_x(t)$, Eq.\eqref{EOM} being replaced by
\begin{equation}
\mathcal{J}_\alpha(t)=J_{\alpha}
+\i\int_0^t[\e^{-\i uH}V\e^{\i uH},\mathcal{J}_\alpha(u)]\d u.
\label{Bcal1}
\end{equation}
Using Eq.\eqref{lesserSS} with $s=0$ in Eq.\eqref{current1} we get
\begin{equation}
\label{current6}
I_{\alpha,\boxdot}=\lim_{t\to\infty}\langle J_{\alpha}(t)\rangle_\boxdot=
\langle\mathcal{J}_\alpha[\omega_+^\ast]\rangle_{L,\boxdot},
\end{equation}
where $\mathcal{J}_\alpha=\lim_{t\to\infty}\mathcal{J}_\alpha(t)$ is calculated by setting $t=\infty$
in the Dyson expansion generated by iteration of Eq.\eqref{Bcal1} and  
$\mathcal{J}_\alpha[\omega_+^\ast]$ is obtained in the usual way from 
$\mathcal{J}_\alpha$.
Comparing the final formulas for the two cases `$\p$' and `$\pf$', one realizes that
$I_{\alpha,\pf}(\vec{v}=\vec{0})=I_{\alpha,\p}(\vec{\mu}=[\mu_0,\!..,\mu_0])$. 
Since $J_\alpha=-\i[H+V,N_\alpha]=-\i[H^{(\Lambda)}+V,N_\alpha^{(\Lambda)}]$, one has
$$
I_{\alpha,\pf}(\vec{v}=\vec{0})=-\lim_{\Lambda\to\infty}
{\rm Tr}_{\mathcal{F}^{(\Lambda)}}
\{\rho^{(\Lambda)}_\pf\i[H^{(\Lambda)}+V,N^{(\Lambda)}_\alpha]\}=0,
$$ 
due to the cyclicity of the trace and the fact that $[\rho^{(\Lambda)}_\pf,H^{(\Lambda)}+V]=0$. 
Thus both currents vanish in the absence of bias. This fact cannot be seen  
from the interacting Meir-Wingreen formula \cite{JWM}. 

The interaction effects can be calculated perturbatively from Eq.\eqref{Bcal1}. For the partitioning 
setting with identical leads having a hopping constant $\tau_L>0$ one finds
\begin{equation}\label{currentLB}
I_{\alpha,\p}= I_{\alpha,\mathrm{LB}}+{\cal O}(\xi^2)+{\cal O}(\xi\tau^6),
\end{equation} 
where  $I_{LB}$ assumes a Landauer-like form
\begin{equation}
\label{currentLBB}
I_{\alpha,\mathrm{LB}}=\sum_\gamma \int_{-2\tau_L}^{2\tau_L}(F_{\alpha}(E)-F_{\gamma}(E))
|{\cal T}^{{\rm MF}}_{\alpha\gamma}(E,\xi)|^2\d E
\end{equation}
with the transmittance \cite{CJM} ${\cal T}^{{\rm MF}}_{\alpha\gamma}(E,\xi)$ 
corresponding to a mean-field Hamiltonian $h_{S,{\rm MF}}=h_S+\xi v_{{\rm MF}}$
where
\begin{equation}
v_{{\rm MF}}=\sum_{m\in S}v_{H,m}|m\rangle\langle m|
-\sum_{m,n\in S} v_{X,mn}|m\rangle\langle n|
\end{equation}
and
$$
v_{H,m}=\sum_{n\in S} v(m,n)\langle n|\rho_+|n\rangle,
\quad
v_{X,mn}=v(m,n)\langle n|\rho_+|m\rangle,
$$ 
are Hartree and exchange terms, with the single-particle density operator $\rho_+$
given by Eq.\eqref{rhos}.

\section{Proofs}\label{proof}

We start by analysing the structure of ${\cal A}_x(t)$, following\cite{Jaksic6}.
Iterating Eq.\eqref{EOM} and using Eq.\eqref{Vintrot}, one obtains an infinite series involving iterated
integrals of the multiple commutators 
$$
[\tilde a_{m_1}^\ast(u_1)\cdots\tilde a_{n_1}(u_1),[\cdots,
[\tilde a_{m_r}^\ast(u_r)\cdots\tilde a_{n_r}(u_r),a^\fast_x]\cdots],
$$
where $\tilde a_m(u)=a(\e^{-\i uh}|m\rangle)$. By repeated use of the algebraic identity
\begin{align*}
&[b_1\cdots b_k,c_1\cdots c_l]\\
&=\sum_{i=1}^k\sum_{j=1}^l(-1)^i\{b_i,c_j\}c_1\cdots c_{j-1}b_1\cdots\cancel{b}_i\cdots b_kc_{j+1}
\cdots c_l,
\end{align*}
and of the canonical anti-commutation relations 
$\{a^\ast(|\phi\rangle),a(|\psi\rangle)\}=\langle \psi|\phi \rangle$, 
$\{a(|\phi\rangle),a(|\psi\rangle)\}=0$ we recast our expansion into the form
\begin{align}\nonumber
\mathcal{A}_x(t)&=a_x
+\sum_{r\geq 1}\xi^r\int_{0\le u_r\le\cdots\le u_1\le t} \d u_1\cdots\d u_r\\
&\sum_{G\in\Gamma_r} C_r^G(u_1,\!..,u_r;x){\cal M}_r^G(u_1,\!..,u_r;x),  
\label{Apol}
\end{align}
where each $\Gamma_r$ is a finite set (of contraction diagrams). For each $G\in\Gamma_r$,
${\cal M}_r^G$ is a monomial of type \eqref{LemmaL1} and $C_r^G$
is a product of `pairing factors' like $\langle y|\e^{-\i u_jh}|y'\rangle $ or 
$\langle y|\e^{-\i(u_j-u_k)h}|y' \rangle $, where $y,y'\in S\cup\{x\}$. 

Our first assumption ensures that there exists a constant $C_x$ such  that
$$
\int_0^\infty\max_{y,y'\in S\cup\{x\}}|\langle y|\e^{-\i uh}|y'\rangle|\,\d u\le C_x.
$$
A delicate combinatorial analysis then shows that (see Theorem 1.1 in \cite{Jaksic6})
$$
\sum_{r\geq 1}|\xi|^r\int_{0\le u_r\le\cdots\le u_1\le\infty}\hskip-50pt\d u_1\cdots\d u_r
\sum_{G\in\Gamma_r} |C_r^G(u_1,\!..,u_r;x)|<\infty,  
$$
provided $|\xi|\le\Lambda_0=2/(27|S|^2C_xv)$, where $|S|$ is the number of sites in the sample $S$
and $v=\max_{n,m\in S}|v(n,m)|$.
Thus, the expansion \eqref{Apol} converges uniformly w.r.t.\;$t\in[0,\infty]$.
In particular, setting $t=\infty$ in Eq.\eqref{Apol} yields a convergent
expansion of $\mathcal{A}_x$ (and taking adjoint gives an expansion for $\mathcal{A}^\ast_x$). 
This proves Theorem \ref{Asymptot}. 

To prove Theorem \ref{MainRes}, we first notice that, according to Eq.\eqref{wellknown},
we get an expansion of
\begin{align*}
\e&^{-\i t(H_L+H_B)}a^\#_x(t-s)\e^{\i t(H_L+H_B)}\\
&=\e^{-\i t(H_L+H_B)}\e^{\i(t-s)H}\mathcal{A}^\#_x(t-s)
\e^{-\i(t-s)H}\e^{\i t(H_L+H_B)},
\end{align*}
by replacing each factor $a^\#(|\psi\rangle)$ of any monomial
$\mathcal{M}_r^G$ in Eq.\eqref{Apol} by 
$a^\#(\e^{-\i t(h_L+h_B)}\e^{\i (t-s)h}|\psi\rangle)$. 
Since Eq. \eqref{omegastar} implies that
$$
\lim_{t\to\infty}a^\#(\e^{-\i t(h_L+h_B)}\e^{\i (t-s)h}|\psi\rangle)
=a^\#(\e^{-\i s(h_L+h_B)}\omega_+^\ast|\psi\rangle),
$$
one has
\begin{align*}
\lim_{t\to\infty}\e^{-\i t(H_L+H_B)}&a^\#_x(t-s)\e^{\i t(H_L+H_B)}\nonumber\\
=&\mathcal{A}^\#_x[\e^{-\i s(h_L+h_B)}\omega_+^\ast],
\end{align*}
and hence, $\mathcal{B}(t)=\e^{-\i t(H_L+H_B)}a_y^\ast(t-s)a^\fast_x(t)\e^{\i t(H_L+H_B)}$
satisfies
\begin{equation}
\lim_{t\to\infty}\mathcal{B}(t)=\mathcal{A}^\ast_y[\e^{-\i s(h_L+h_B)}\omega_+^\ast]
\mathcal{A}^\fast_x[\omega_+^\ast].
\label{centraleq}
\end{equation}
Notice that since the range of $\omega_+^\ast$ is $\mathcal{H}_L$, the RHS of this identity 
is an observable of the leads. In the `$\p$' case, $H_B=0$ and the state $\langle\,\cdot\,\rangle_\p$
is invariant under the dynamics of $H_L$. It follows that
\begin{align*}
\lim_{t\to\infty}\langle a_y^\ast(t-s)a_x(t)\rangle_\p
&=\lim_{t\to\infty}\langle\mathcal{B}(t)\rangle_\p\\
&=\langle\mathcal{A}^\ast_y[\e^{-\i sh_L}\omega_+^\ast]
\mathcal{A}^\fast_x[\omega_+^\ast]\rangle_\p\\
&=\langle\mathcal{A}^\ast_y[\e^{-\i sh_L}\omega_+^\ast]
\mathcal{A}^\fast_x[\omega_+^\ast]\rangle_{L,\p},
\end{align*}
which proves Theorem \ref{MainRes} in the `$\p$' case.

To deal with the `$\pf$' case, we invoke standard perturbation theory (see e.g. \cite{Kubo})
to write
\begin{equation}
\langle\mathcal{O}\rangle_\pf
=\frac{\langle\mathcal{D}\mathcal{O}\rangle_\d}{\langle\mathcal{D}\rangle_\d}.
\label{Kubo}
\end{equation}
There, $\langle\,\cdot\,\rangle_\d$ denotes the grand canonical ensemble for
the decoupled dynamics $H_0+V$ at inverse temperature $\beta$ and
chemical potential $\mu_0$, {\sl i.e.,} the product state
$$
\langle\mathcal{O}_L\mathcal{O}_S\rangle_\d
=\langle\mathcal{O}_L\rangle_{L,\pf}\,
\frac{\mathrm{Tr}_{\mathcal{F}_S}\{\e^{-\beta(H_S+V-\mu_0N_S)}\mathcal{O}_S\}}
{\mathrm{Tr}_{\mathcal{F}_S}\{\e^{-\beta(H_S+V-\mu_0N_S)}\}},
$$
and
\begin{align*}
{\cal D}&=\e^{\beta(H_0+V-\mu_0 N)}\e^{-\beta(H+V-\mu_0 N)}\\
&=I+\sum_{k\geq 1}(-1)^k\int_0^{\beta}\d\tau_1\cdots\int_0^{\tau_{k-1}}\d\tau_k
{\hat H}_T(\tau_1)..{\hat H}_T(\tau_k),
\end{align*}
where ${\hat H}_T(u)=\e^{u (H_0+V)}H_T\e^{-u (H_0+V)}$. The state $\langle\,\cdot\,\rangle_\d$
being invariant under the dynamics of $H_L+H_B$, one has
$$
\langle\mathcal{D}a_y^\ast(t-s)a_x(t)\rangle_\d
=\langle\mathcal{D}_t\mathcal{B}(t)\rangle_\d,
$$
where $\mathcal{D}_t=\e^{-\i t(H_L+H_B)}\mathcal{D}\e^{\i t(H_L+H_B)}$. It follows
from Eq.\eqref{centraleq} that
\begin{align}
\lim_{t\to\infty}&\langle\mathcal{D}a_y^\ast(t-s)a_x(t)\rangle_\d\nonumber\\
&=\lim_{t\to\infty}\langle\mathcal{D}_t
\mathcal{A}^\ast_y[\e^{-\i s(h_L+h_B)}\omega_+^\ast]
\mathcal{A}^\fast_x[\omega_+^\ast]\rangle_\d.
\label{DAlim}
\end{align}
Introducing the partial trace
$$
\mathcal{D}_L=\frac{\mathrm{Tr}_{\mathcal{F}_S}\{\e^{-\beta(H_S+V-\mu_0N_S)}\mathcal{D}\}}
{\mathrm{Tr}_{\mathcal{F}_S}\{\e^{-\beta(H_S+V-\mu_0N_S)}\}},
$$
we observe that, for any observable $\mathcal{O}_L$ of the leads,
$$
\langle\mathcal{D}_t\mathcal{O}_L\rangle_\d
=\langle\e^{-\i t(H_L+H_B)}\mathcal{D}_L\e^{\i t(H_L+H_B)}\mathcal{O}_L\rangle_{L,\pf}.
$$
Since $h_L+h_B$ has continuous spectrum, the dynamics of $H_L+H_B$ is 
mixing w.r.t.\;the state $\langle\,\cdot\,\rangle_{L,\pf}$ (see e.g.\cite{Jaksic4}) so that
\begin{align*}
\lim_{t\to\infty}\langle\mathcal{D}_t\mathcal{O}_L\rangle_\d
&=\lim_{t\to\infty}\langle\e^{-\i t(H_L+H_B)}\mathcal{D}_L\e^{\i t(H_L+H_B)}
\mathcal{O}_L\rangle_{L,\pf}\\
&=\langle\mathcal{D}_L\rangle_{L,\pf}\langle\mathcal{O}_L\rangle_{L,\pf}
=\langle\mathcal{D}\rangle_{\d}\langle\mathcal{O}_L\rangle_{L,\pf}.
\end{align*}
Applying this identity to the RHS of Eq.\eqref{DAlim} and inserting the result into
Eq.\eqref{Kubo} proves Theorem \ref{MainRes} in the `$\pf$' case.

\section{ Conclusions}  

We give sufficient conditions for the existence of a steady-state regime 
for open interacting systems, using an approach based on time-dependent scattering theory.  
While the approach is perturbative 
w.r.t.\;the interaction strength, the existence of the stationary regime is general in the 
sense that it does not rely on a particular approximation scheme for the Coulomb effects. 
To our best knowledge, the steady-state regime is not proved within the Keldsyh formalism: 
the steady-state value of the current is derived by {\it assuming} that the two-time GFs depend 
only on time difference \cite{JWM}. We perform the thermodynamic limit 
without making use of Langreth rules and Dyson equations. In the Keldysh approach 
it is not clear how to perform the thermodynamic limit on the {\it proper} interaction self-energy. 

 The smallness condition on $\xi$ for the existence of the
 steady-state is also necessary. 
For example, if $v(m,n)$ is diagonal then $V$ is a single-particle operator (Hartree-like potential) 
which can create bound states for large values of $\xi$, leading thus to oscillations similar to the
ones observed in \cite{Stefanucci}. However, we conjecture that if one performs 
the ergodic limit $\lim_{T\to\infty}\frac{1}{T}\int_0^{T}J_{\alpha}(t)\d t$, a time-averaged 
steady-state could still be achieved even in the general interacting case. It would be 
interesting to investigate the ergodic limit of the oscillating currents reported 
in Fig.\,1 from \cite{Kurth2}, which seem to support our conjecture.    

 In the partitioning approach we have shown that the steady-state
quantities {\it do not depend on the initial many-body configuration
  of the sample $\rho_{S,\p}$}. 
Moreover, one can allow other switching procedures of the bias or of the lead-sample 
coupling (not just the sudden one), and the steady state remains unchanged (the complete 
proof will be given in Ref. \onlinecite{CMP}). Let us mention that very recently\cite{CM} 
we have shown that when $\xi$ is allowed to be arbitrarily large and the system is in the 
off-resonant regime in which $h$ has eigenvalues situated very far from the continuous spectrum, 
the {\it ergodic} cotunneling current {\it presents memory effects and
  depends on $\rho_{S,\p}$ if $\xi\neq 0$}. 
In the non-interacting case, we still have independence on $\rho_{S,\p}$. 


Our results could be numerically implemented in 
both settings (partitioning and partition-free) and compared to the ones obtained from the Keldysh 
formalism. The second correction in Eq.\eqref{currentLB} suggests significant differences for strong 
coupling to the leads.

\begin{acknowledgments}
V.\,M.\ acknowledges the financial support from PNCDI2 program (grant No.\ 515/2009), 
Core Project (grant No.\ 45N/2009) and the Danish FNU Grant {\it Mathematical Physics}.
The research of C.-A.\,P. was partly supported by ANR (grant 09-BLAN-0098).
\end{acknowledgments}

\bibliographystyle{apsrev}

\end{document}